\documentclass[pra,superscriptaddress,amsmath,amssymb,twocolumn]{revtex4-2}

\usepackage[usenames,dvipsnames]{xcolor}
\usepackage{amsmath,amssymb}
\usepackage{graphicx}
\usepackage{hyperref}
\usepackage{braket}
\usepackage[normalem]{ulem}
\usepackage{enumerate}
\usepackage{MnSymbol}
\usepackage{esint}
\usepackage{comment}

\newcommand{\be}{\begin{equation}}
\newcommand{\ee}{\end{equation}}
\def\bes{\begin{subequations}}
\def\esu{\end{subequations}}

\newcommand{\dr}{\text{dr}}

\newcommand{\rhoq}{\rho_\text{q}}

\newcommand{\com}[1]{{}} 

\newcommand{\dd}{{\rm d}}

\begin{document}

\newcommand{\titleinfo}{Observation of a generalized Gibbs ensemble in photonics}

\title{\titleinfo}

\author{Alvise Bastianello}
\thanks{These authors contributed equally to this work.}
\affiliation{Technical University of Munich, TUM School of Natural Sciences, Physics Department, 85748 Garching, Germany}
\affiliation{Munich Center for Quantum Science and Technology (MCQST), Schellingstr. 4, 80799 M{\"u}nchen, Germany}

\author{Alexey Tikan}
\thanks{These authors contributed equally to this work.}
\affiliation{Laboratoire Temps-Fr\'equence, Universit\'e de Neuchâtel, Avenue de Bellevaux 51, Neuchâtel, Switzerland}

\author{Francois Copie}
\affiliation{Univ. Lille, CNRS, UMR 8523 - PhLAM - Physique des Lasers Atomes et Mol\'ecules, F-59 000 Lille, France}
\author{Stephane Randoux}
\affiliation{Univ. Lille, CNRS, UMR 8523 - PhLAM - Physique des Lasers Atomes et Mol\'ecules, F-59 000 Lille, France}

\author{Pierre Suret}
\email{pierre.suret@univ-lille.fr}
\affiliation{Univ. Lille, CNRS, UMR 8523 - PhLAM - Physique des Lasers Atomes et Mol\'ecules, F-59 000 Lille, France}

\begin{abstract}
In generic classical and quantum many-body systems, where typically energy and particle number are the only conserved
quantities, stationary states are described by thermal equilibrium. In contrast, integrable systems showcase an infinite hierarchy of conserved quantities that inhibits conventional thermalization, forcing relaxation to a
Generalized Gibbs Ensemble (GGE) -- a concept first introduced in quantum many-body physics. In this study, we provide experimental evidence for the emergence of a GGE in a photonic system.
By investigating partially coherent waves propagating in a normal
dispersion optical fiber, governed by the one‐dimensional defocusing nonlinear Schroedinger equation, we directly measure the density of states of the spectral parameter (rapidity) to confirm the time invariance of the full set of conserved charges. We also observe the relaxation of optical power statistics to the GGE’s theoretical prediction, obtained using the experimentally measured density of states. These complementary measurements unambiguously establish the formation of a GGE in our photonic platform, highlighting its potential as a powerful tool for probing many-body integrability and bridging classical and quantum integrable systems.
\end{abstract}

\maketitle

\section{Introduction}
\label{sec_intro}
Strongly nonlinear systems and field theories out of equilibrium are among the most arduous challenges of modern physics. 
While determining every aspect of time evolution is a daunting problem, large scales see the emergence of universality and thermodynamics, dictating relaxation to the maximally entropic state compatible with conservation laws \cite{Jaynes1957,Jaynes1957B}.
When only the Hamiltonian is conserved, the system relaxes to a Gibbs thermal state.
However, there are several known exceptions to this paradigm \cite{Abanin2019,Serbyn2021,Gromov2024}, such as integrable systems.
Integrable systems are strongly interacting one-dimensional models with an infinite number of extensive conservation laws that constrain the dynamics and defy naive thermodynamics. 
Cutting-edge experiments \cite{Bloch2012,Guan2022,Suret2024} spurred two very active but seemingly disconnected communities in tremendously advancing understanding nonequilibrium, the first focusing on quantum systems \cite{Smirnov1992, Korepin1993} and the other on classical partial differential equations (PDEs) \cite{Faddeev1987}.
Since emergent statistical properties are determined by symmetries and conservation laws, both communities independently rediscovered similar concepts from different angles. 

The quantum community focused on \emph{quantum quenches} \cite{Polkovinkov2011}, simple initial-value problems where a quantum state is unitarity evolved. 
Under minimal assumptions \cite{Sotiriadis2014,Doyon2017}, at late times an integrable system locally relaxes to a Generalized Gibbs Ensemble (GGE), first conjectured by Rigol \emph{et al.} \cite{Rigol2007} in 2007. 
The GGE state $\hat{\rho}\propto e^{-\sum_n \mu_n \hat{Q}_n}$ generalizes conventional thermal Gibbs ensemble to the infinite number of conserved quantities, or charges, of integrable systems $\partial_t\hat{\mathcal{Q}}_n=0$. The coefficients $\mu_n$ are determined by imposing charge conservation between the GGE and the initial conditions.
\begin{figure}[b!]
\centering
\includegraphics[width=0.99\columnwidth]{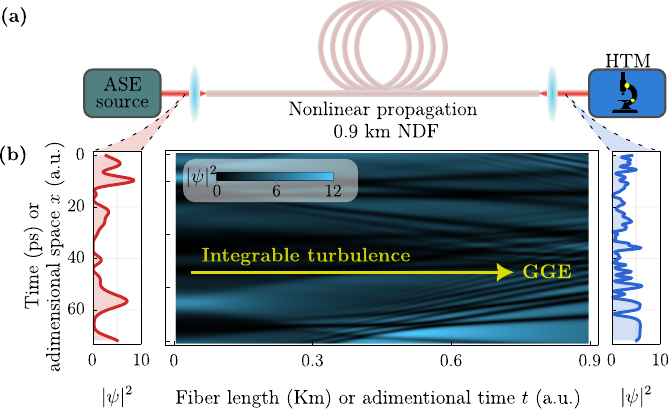}
\caption{\textbf{The GGE from integrable dynamics in optical fibers.---}
Panel (a): Experimental realization of the NLS through light propagation in nonlinear optical fibers.
Panel (b): Evolution of a representative field configuration in the strongly nonlinear regime. Physical time and fiber lengths in the experiment are respectively proportional to space and time in the adimensional NLS equation \eqref{eq_nls}.
}	\label{fig_sketch}
\end{figure}
The GGE concept opened to several groundbreaking theoretical achievements \cite{Essler2011,Kormos2014,Pozsgay2014,Essler2016, Ilievski2016B, Calabrese2016}, and the GGE signature in correlation functions has been experimentally verified in cold atoms \cite{Langen2015}.
Experimental refinements further characterized GGEs, by recovering the root density or density of states \cite{takahashi2005thermodynamics}--equivalent to measuring all the conserved charges and thus uniquely identifying the GGE~\cite{Ilievski2016}--from atomic cloud expansions \cite{Wilson2020,Malvania2021,Dubois2024,Le2023,Li2023,horvath2025}.

In parallel, Zakharov introduced the concept of integrable turbulence for PDEs in 2009, describing the evolution of random initial conditions in integrable systems such as the 1D nonlinear Schroedinger (NLS) or Korteweg–De Vries equations~\cite{Zakharov2009}. This seminal work has since inspired numerous theoretical studies~\cite{Agafontsev2015, gelash2019bound, Congy2024} as well as experiments in optics~\cite{tikan2018SingleshotMeasurementPhase, kraych2019statistical} and hydrodynamics~\cite{Michel2020, Redor:21,suret2020NonlinearSpectralSynthesis}. Most experimental investigations have focused on statistical measurements, particularly the single-point probability density function (PDF) of the intensity \cite{kraych2019statistical, tikan2018SingleshotMeasurementPhase, Randoux2014,Randoux2016Nonlinear}. 
Numerical simulations and experiments have shown that, through integrable turbulence, the system eventually relaxes toward a statistically stationary state \cite{Zakharov2009, Agafontsev2015, kraych2019statistical}.
The natural assumption is relaxation to a GGE determined by the density of states in analogy with the quantum framework.

In this work, we present the experimental observation of relaxation to a GGE following a ``quench'' in a classical system, realized through the propagation of light in a single-mode optical fiber from random initial conditions as illustrated in Fig.~\ref{fig_sketch}(a).
Light propagation in the laboratory frame is captured by the integrable defocusing NLS equation~\cite{agrawal2013NonlinearFiberOptics} $
i2\partial_z A=  \beta_2\partial^2_T A-2\gamma |A|^2 A$,
where $A$ is the slowly varying envelope of the optical field, whereas $\beta_2$ and $\gamma$ are the group velocity dispersion and Kerr nonlinearity of the fiber and $z$ and $T$ indicate the propagation distance along the fiber and the moving time frame.
Through this work, we use adimensional units and write the NLS equation in the canonical form
\be
i \partial_t \psi =-\partial^2_x \psi+2|\psi|^2\psi ,
\label{eq_nls}
\ee
where $\psi= A/ \sqrt{P_0}$ and $P_0$ is the average optical power. The adimensional time $t$ and position $x$ are connected with the laboratory coordinates as $t=-\gamma P_0 z/2$, and $x=\sqrt{\gamma P_0/|\beta_2|}\, T$, see Appendix \ref{SI:experiment} for details. The GGE emerges as a statistical description of the field after a long time $t$ equivalent to propagation through a long fiber, see Fig. \ref{fig_sketch}(b).

Previous works experimentally \cite{Bromberg2010} and theoretically \cite{Derevyanko2012} studied the intensity's PDF obtained obtained from initial configurations within a compact support, propagated after long times to distances much larger than its initial width. In contrast, the GGE emerges when the thermodynamic limit is taken first, and the long-time limit afterwards.
We extract the density of states directly from the experimental data before and after evolution, demonstrating its conservation in time, and show it captures the emergent GGE by focusing on the PDF of the intensity.
The intensity's PDF $P(|\psi|^2)$ is the probability of measuring a certain value of $|\psi(x)|^2$ in a translationally invariant ensemble of fields: we compare the profiles evolved from the initial random field configurations with recent theoretical predictions on GGEs \cite{DelVecchio2020}.
This measurement is challenging because probing the density of states requires an accurate single-shot recording of both phase and amplitude of the field.
Our work places experimental platforms for classical integrable PDEs as a promising venue for investigations historically pertaining to the quantum realm, and poses a crucial milestone in establishing a fruitful dialogue between the communities working on quantum and classical many-body integrability, recently initiated with the parallelism \cite{Bonnemain2022} between soliton gases \cite{Zakharov1971,El2005,El2021,Suret2024} and the more recent Generalized Hydrodynamics \cite{Doyon2016,Bertini2016}, further strengthened by semiclassical limits of quantum integrability \cite{DeLuca2016,Bastianello2018C,DelVecchio2020,Koch2022,Koch2023,Bezzaz2023,Bastianello2024}.

\section{The experimental setup}
\label{sec_exp}

\begin{figure*}[t!]
\centering\includegraphics[width=0.99\textwidth]{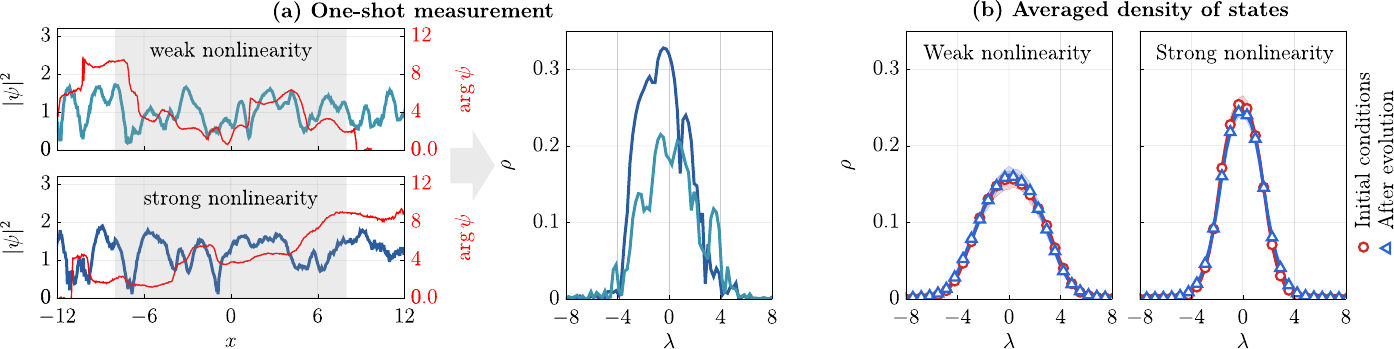}
\caption{\textbf{Extracting the GGE from experimental data.---}
Panel $(a)$: We show the profile of the norm and phase of one random field configuration after the weak and strong nonlinear evolution. To avoid boundary detection problems, we consider a central window (shadowed area $x\in[-8,8]$) and compute the density of states within it, shown in the rightmost panel in the respective colors. Panel $(b):$ We show the conservation of the density of states by reporting its average values before and after the evolution, for weak and strong nonlinearities. When averaging, we consider $10^3$ experimental samples: the confidence interval (shaded areas) is obtained by considering the maximum spreading of partial averages done on $250$ samples each.
}
	\label{fig_exp}
\end{figure*}
The conceptual scheme of the experimental setup is illustrated in Fig.~\ref{fig_sketch} (a).
The initial random conditions are prepared by shaping the spectrum of the amplified spontaneous emission (ASE) signal (Highwave Optical Technologies operating at central wavelength 1550 nm) with a programmable optical filter to a Gaussian profile. The optical power after a post-amplification stage has been adjusted with a variable attenuator.
As a result, the initial condition corresponds to the superposition of independently and gaussianly distributed Fourier modes having random phase. This initial field distribution gives an exponential PDF of the optical power~\cite{Agafontsev2015}.
The random light is injected in 900 m-long single mode fiber with a group velocity dispersion coefficient $\beta_2 = 22$ ps$^2$/km and a nonlinear Kerr coefficient $\gamma = 3$ W$^{-1}$km$^{-1}$.
Our study requires the full field detection of non-periodic ultrafast optical signals, typically in the sub-picosecond regime. For this purpose, we employed a heterodyne time microscope (HTM)~\cite{tikan2018SingleshotMeasurementPhase}, which leverages space-time imaging analogies~\cite{suret2016SingleshotObservationOpticala}. The HTM enables single-shot recording of both phase and power profiles of the optical signals with a temporal resolution of 250 fs over a temporal window of 40 ps. The details of the data acquisition process with HTM are provided in the Appendix \ref{SI:experiment} as well as in Refs.~\cite{tikan2018SingleshotMeasurementPhase,lebel2021SingleshotObservationBreathers}.

The effective strength of the nonlinearity is tuned by controlling the adimensional momentum width $\Delta k=2\pi\sqrt{\beta_2/(\gamma P_0)} \,\Delta \nu$, where $\Delta \nu$ is the optical bandwidth (in physical unit, Hz) of the initial signal. For large $\Delta k$, the kinetic (linear) term in the NLS equation \eqref{eq_nls} is dominant and the nonlinearity is weak. For small $\Delta k$, the nonlinear interaction gains importance. 
In this study, we used $\Delta \nu = 0.5$ THz and $P_0=3.5$ W for the weakly nonlinear case, and $\Delta \nu = 0.1$ THz and $P_0=3.6$ W for the strongly nonlinear case which corresponds to $\Delta k =2.73$ and $\Delta k=0.539$ for the normalized momentum, respectively, see Appendix \ref{SI:experiment}.

\section{The density of states}
\label{sec_dos}

In translationally invariant and linear PDEs, normal modes are the Fourier components and their spectrum --which is conserved in time-- is the density of states. In nonlinear, or interacting, integrable models normal modes become highly nonlinear functions of the field and the density of states is not the Fourier spectrum any longer. Instead of a simple Fourier transform, identifying normal modes requires the Inverse Scattering Transform (IST) \cite{Faddeev1987} in classical PDEs, and the Bethe Ansatz \cite{Korepin1993} in quantum integrability. Here we briefly overview standard concepts of non-linear modes in integrable models and their density of states, and leave to the Appendices \ref{sec_qTBA} and \ref{sec_clTBA} a detailed and pedagogical overview in the quantum and classical cases respectively.
Nonlinear modes in classical PDEs are generally radiation and solitons \cite{Faddeev1987}, whereas fermionic excitations emerge \cite{takahashi2005thermodynamics} in the quantum case.
In both cases, modes are characterized by a spectral parameter $\lambda$ which in the quantum case, and through this work, is called the rapidity. 
These modes behave as ballistic quasiparticles showcasing only factorized elastic scattering \cite{Dorey1997}, completely determined by the two-body scattering shift. 
In practice, their dynamics is reminiscent of a hard rods gas \cite{Boldrighini1983,Doyon2017HR}, i.e. extended particles with elastic pairwise scattering, where the effective length depends on the rapidities of the colliding particles \cite{Doyon2018}.
Each mode additively contributes to conserved charges: in the thermodynamic limit, one conveniently introduces a coarse grain mode density $\rho(\lambda)$ as $\mathcal{Q}_n=L\int \dd\lambda\, q_n(\lambda)\rho(\lambda)$, where $L$ is the large system size and $q_n(\lambda)$ the single-mode contribution to the charge. Since $\rho(\lambda)$ it parametrizes the charges' densities, which are thermodynamically-intensive quantities, it is self-averaging. In the classical and quantum cases, $\rho(\lambda)$ is commonly known as the density of states \cite{El2005} or root density \cite{takahashi2005thermodynamics} respectively. Since $\rho(\lambda)$ determines all the charges, it is conserved in time and unambiguously fixes the GGE \cite{Ilievski2015,Ilievski2016}.
In classical PDEs, the density of states is retrievable from the IST spectrum of field's configurations \cite{yang2010nonlinear}: hereon, we consider the defocusing NLS. 
In this case, the spectrum is encompassed by a single radiative mode with real rapidity $\lambda$, whose density of states is computable as follows \cite{DeLuca2016,DelVecchio2020,Bezzaz2023}.
We focus on smooth field configurations in a large box of volume $L$ centered in zero and introduce the auxiliary linear problem \cite{Faddeev1987} $\partial_x\Psi(x)=U(x)\Psi(x)$, where $\Psi(x)$ is an auxiliary 2-component complex field. The $2\times 2$ matrix $U(x)=\Re[\psi(x)]\sigma^x+\Im[\psi(x)]\sigma^y+\frac{\lambda}{2i}\sigma^z$ is the parallel transporter and $\sigma^{x,y,z}$ are the canonical Pauli matrices. 
The auxiliary problem transforms the initial condition $\Psi(-L/2)={\footnotesize\begin{pmatrix}1\\0\end{pmatrix}}$ into $\Psi(L/2)={\footnotesize\begin{pmatrix}e^{-i x \lambda}a(\lambda)\\b(\lambda)\end{pmatrix}}$, whose first component $a(\lambda)$ generates the conservation laws $\mathcal{Q}_n$\cite{Faddeev1987} and parametrizes the density of states \cite{DeLuca2016,DelVecchio2020,Bezzaz2023} $\rho(\lambda)=\lim_{L\to\infty} \frac{1}{\pi L} \log|a(\lambda)|$, see also Appendix \ref{sec_clTBA}.
The density of states $\rho(\lambda)$ is then obtained tabulating the solution of the auxiliary problem for different values of $\lambda$.
In the limit where the nonlinearity in Eq. \eqref{eq_nls} is negligible, the density of states approaches the Fourier spectrum and the rapidity becomes the Fourier wavevector, but they differ at strong nonlinearities.
In Fig. \ref{fig_exp}, we show the density of states extracted from the experiment, both for the weakly nonlinear and strongly nonlinear regimes, and verify its conservation during time evolution. The density of states is extracted by discretizing and then numerically solving the auxiliary linear problem for different values of $\lambda$. See Appendix \ref{sec_clTBA} for details on the numerical implementation and \cite{Zenodo} for a commented code.
Arbitrarily chosen examples of the recorded power and phase profiles after the propagation in the fiber are shown in Fig.~\ref{fig_exp}a.
The nonlinear evolution exhibits rich dynamics including dispersive shock waves~\cite{el2016DispersiveShockWaves} and dark soliton~\cite{kivshar1998DarkOpticalSolitons} formation, as shown in spatiotemporal diagram for similar conditions reported in Fig.~\ref{fig_sketch}(b).
Interestingly, dark solitons do not explicitly contribute to the GGE, which is completely determined by the radiative mode only \cite{Bullough1986,DelVecchio2020}.
However, kinetic equations for a dark solitons gas in the defocusing NLS have been derived in the framework of finite gap theory~\cite{tovbis2025}, raising interesting questions on the relation of the latter with the discussed GGE.
The corresponding single-shot density of states are plotted in Fig.~\ref{fig_exp}a (right panel) exhibiting a substantial shot-to-shot variation.
To probe simultaneously conservation laws, we verify the time-independence of the density of states extracted from the experimental data. 
Field configurations at the initial and final time have been recorded for independent realizations, allowing us to verify the conservation of the density of states only on average.
We analyzed 1000 data frames for both the initial conditions and the signal after the propagation in the fiber, restricting the confidence interval to $\approx$26 ps window in the center of the frame, corresponding to $x\in[-8,8]$ in rescaled units, to minimize the influence of reduced signal-to-noise ratio at the edges of the frame.

Importantly, fig.~\ref{fig_exp}b demonstrates an excellent correspondence between the initial and final density of states $\rho(\lambda)$, confirming the conservation of charges and thus that we operate in the integrable regime, showing corrections to Eq. \eqref{eq_nls} coming from experimental imperfections (including losses) are negligible.
The bandwidth of the initial Fourier modes determines the width of the density of states and the nonlinearity's strength. 
For weak interactions, the density of states is close to the momentum distribution and the two have similar width, whereas the two significantly depart in the strongly nonlinear regime.

\section{Emergence of the GGE}
\label{sec_gge}

\begin{figure}[t!]
\centering
	\includegraphics[width=0.99\columnwidth]{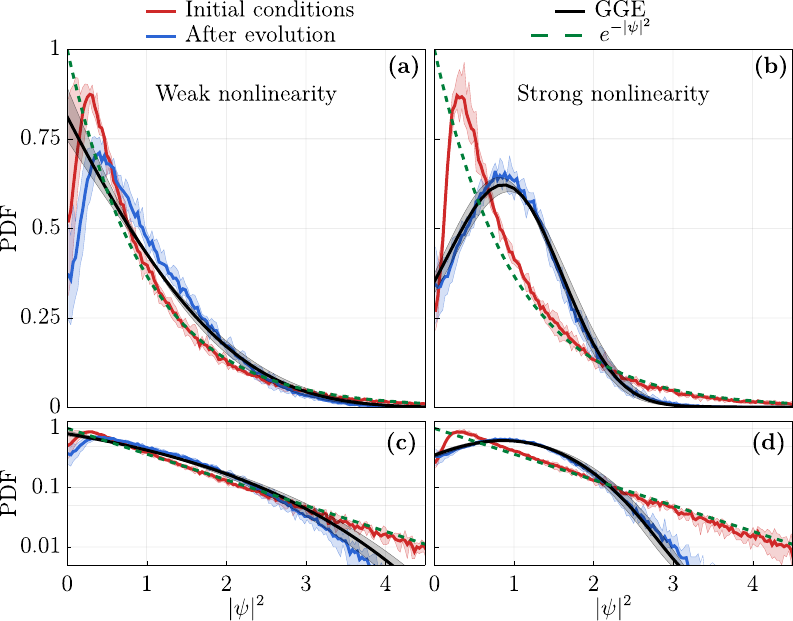}
\caption{\textbf{Relaxation of the intensity PDF to the GGE.---} We show the PDF of the intensity field $|\psi|^2$ extracted from the initial data on the initial conditions and after evolution (red line and blue line respectively), in linear \textbf{(a,b)} and log \textbf{(c,d)} scale. We compare PDF data with the GGE result (black line) \eqref{eq_GGE_PDF} obtained from the density of states extracted from experimental data, see Fig. \ref{fig_exp}. We compare with the exponential decay from Gaussian independent plane waves as reference (dashed line).  As before, we show the average over $10^3$ independent samples, the confidence interval is the maximum spreading from partial averages of 250 samples each. The experimental PDF is extracted from the same window of data $x\in[-8,8]$ used for the density of states in  Fig. \ref{fig_exp}, and the bin size is 0.03. }
	\label{fig_Exp_comparison}
\end{figure}

While the density of states remains unaffected through the dynamics, other observables showcase non-trivial evolution and eventual relaxation to a stationary state described by the GGE. For example, the PDF of the normalized power is initially an exponential $P_0(|\psi|^2)=e^{-|\psi|^2}$ as expectd for the ensemble of independent Gaussian plane waves. At late times, under the effect of the non-linear evolution, it relaxes to a new distribution determined by the GGE $P_\text{GGE}$.
Our key result is showing the experimentally measured power's PDF relaxes to the corresponding theoretical prediction for GGEs derived in Ref. \cite{DelVecchio2020}, using semiclassical limits \cite{DeLuca2016} of previous results in quantum integrability \cite{Bastianello2018,Bastianello2018B}.
The PDF of the intensity on a GGE has the integral representation
\be\label{eq_GGE_PDF}
P_\text{GGE}(|\psi|^2)=16\int_0^\infty \dd \theta J_0(8\sqrt{\theta |\psi|^2})e^{-\frac{16}{\pi}\int_0^{\sqrt{\theta}}\dd\tau\, \tau S(\tau)}\, ,
\ee
where $J_0(x)$  is a modified Bessel function of the first kind, $S(\tau)\equiv \int \dd\lambda\, s_{\tau}(\lambda)$, and $s_{\tau}(\lambda)$ solves the auxiliary integral equation
\be\label{eq_s}
[\vartheta^{-1}(\lambda)+i0^+]s_\tau(\lambda)=1+\fint \frac{\dd\lambda'}{2\pi} \frac{2}{\lambda-\lambda'}(4i\tau-\partial_{\lambda'})s_\tau(\lambda')\, .
\ee
Above, the singular integral is meant within a principal value regularization, and we introduce other standard quantities from quantum integrability \cite{takahashi2005thermodynamics} as the filling fraction $\vartheta(\lambda)=\rho(\lambda)/\rho^t(\lambda)$, and the total root density $\rho^t(\lambda)=\tfrac{1}{2\pi}-\fint\frac{\dd\lambda'}{2\pi}\tfrac{2}{\lambda-\lambda'}\partial_{\lambda'}\rho(\lambda')$. The regularization $i0^+$ is needed for Eq. \eqref{eq_s} to be well-defined, and it results in zeroes in $e^{-\frac{16}{\pi}\int_0^{\sqrt{\theta}}\dd\tau S(\tau)}$: details on the practical computations of the PDF are provided in Appendix \ref{sec_qtocl}, see also Zenodo \cite{Zenodo} for a working code. As we show in the Appendix \ref{sec_qtocl}, Eq. \eqref{eq_GGE_PDF} recovers the expected exponential distribution for weak non-linearities, but it sensibly departs from it at stronger non-linearities.
In Fig.~\ref{fig_Exp_comparison}, we compare experimental measurements of the PDF with theoretical predictions, having the experimental density of states values (Fig~\ref{fig_exp}b) as the input. 
The PDF was computed on the same 1000 independent field configurations previously used in Fig~\ref{fig_exp}b, and considering spatial averaging within the same central window. The confidence interval is estimated from the maximum spreading of partial averages on four groups of 250 samples each. The theory uncertainty is analogously estimated: we extract the density of states for each group of 250 samples, compute the relative PDF and consider the maximum spreading. The PDF solid line is obtained from the density of states average over all field configurations.
Fig.~\ref{fig_Exp_comparison}a depicts the weakly nonlinear case, where the PDF of the stationary state only slightly deviates from the initial exponential distribution (red line). 
The discrepancy between the experimental data and the prediction at low power levels is mainly attributed to the finite resolution of the HTM. Fig.~\ref{fig_Exp_comparison}b illustrates the strongly nonlinear case, where the PDF undergoes a substantial transformation due to nonlinear propagation, with a non-monotonous behavior.
Also in this regime, experimental data agree well with GGE theorerical predictions. 
Altogether, the combined measurements provided in Fig. \ref{fig_exp} and Fig. \ref{fig_Exp_comparison} show the emergence of a GGE stationary state, determined by the initial conditions through the conservation laws.

\section{Discussion}
\label{sec_discussion}

Photonic devices, particularly optical fibers, serve as precise and versatile platforms for studying nonlinear waves. Our results demonstrate that photonics can address fundamental questions traditionally explored in quantum many-body physics. Specifically, we provide a full characterization of the classical analog of a ``quantum quench" \cite{Polkovinkov2011} within the integrable classical defocusing NLS equation.
We investigated the emergence of a GGE from a random-wave initial condition by measuring the density of states, analogous to the rapidity distribution in quantum systems, and confirming the relaxation of the experimental PDF  of $|\psi|^2$ to the theoretical GGE prediction \cite{DelVecchio2020}. The characterization of the dynamics of the quantum counterpart of the NLS equation, the Lieb-Liniger model \cite{Lieb1963A,Lieb1963B}, attracted considerable experimental effort in cold atoms \cite{Kinoshita2006}. The density of states, or rapidity distribution as it is often denoted in these experiments, is indirectly measured through atomic cloud expansions \cite{Wilson2020,Malvania2021,Dubois2024,Le2023,Li2023,horvath2025}, and fluctuations of the number of particles \cite{Esteve2006,Armijo2010}, i.e. the quantum analog of the PDF of $|\psi|^2$, can be measured with a finite-interval resolution.
In contrast, photonic platforms enable direct access to the density of states and high-resolution power fluctuation measurements via local detection of the complex field.
These high-precision measurements of the field enable a complete characterization of the state, whereas ultracold gases have access only to limited information: depending on the experimental platform, either coarse grain imaging of $|\psi(x)|^2$ or momentum distributions. Importantly, the destructive measurement process in quantum systems prevents the observation of two-time correlation functions, which can instead be accessed in photonics and have a key-role in the study of integrable systems \cite{Doyon2018C,DeNardis2022}.

Notably, theoretical predictions for fluctuations  of $|\psi|^2$ \cite{DelVecchio2020} were not obtained directly in the NLS, but through semiclassical limits of results in Lieb-Liniger \cite{Bastianello2018,Bastianello2018B}. The study of semiclassical limits in quantum integrability has recently gained considerable momentum \cite{DeLuca2016,Bastianello2018C,DelVecchio2020,Koch2022,Koch2023,Bezzaz2023,Bastianello2024} in addressing unresolved classical problems. Our work elevates this quantum-classical correspondence from theory to experiments, establishing photonics as a powerful tool to complement and potentially surpass cold atom systems in probing the statistical mechanics and hydrodynamics of integrable models.

\section*{Acknowledgements}

AB acknowledges support from the Deutsche Forschungsgemeinschaft (DFG, German Research Foundation) under Germany’s Excellence Strategy–EXC–2111–390814868. AB is grateful to G. Del Vecchio Del Vecchio, A. De Luca and G. Mussardo for collaboration in Ref. \cite{DelVecchio2020}. 
We are grateful to Jerome Dubail for useful comments on the manuscript.
FC, SR and PS acknowledge support by the Agence Nationale de la Recherche  through the SOGOOD (ANR-21-CE30-0061) project, the LABEX CEMPI project (ANR-11-LABX-0007), the Ministry of Higher Education and Research, Hauts de France council and European Regional Development Fund (ERDF) through the Nord-Pas de Calais Regional Research Council and the European Regional Development Fund (ERDF) through the Contrat de Projets Etat-R\'egion (CPER Photonics for Society P4S). FC, SR and PS thank the Centre d'Etudes et de Recherche Lasers et Application (CERLA) for technical support.

\section*{Data and informations availability.} Data, data analysis, and simulation codes are available upon reasonable request on Zenodo~\cite{Zenodo}.

\appendix

\section{Experimental data analysis}
\label{SI:experiment}

\begin{figure}[t!]
\centering
	\includegraphics[width=0.99\columnwidth]{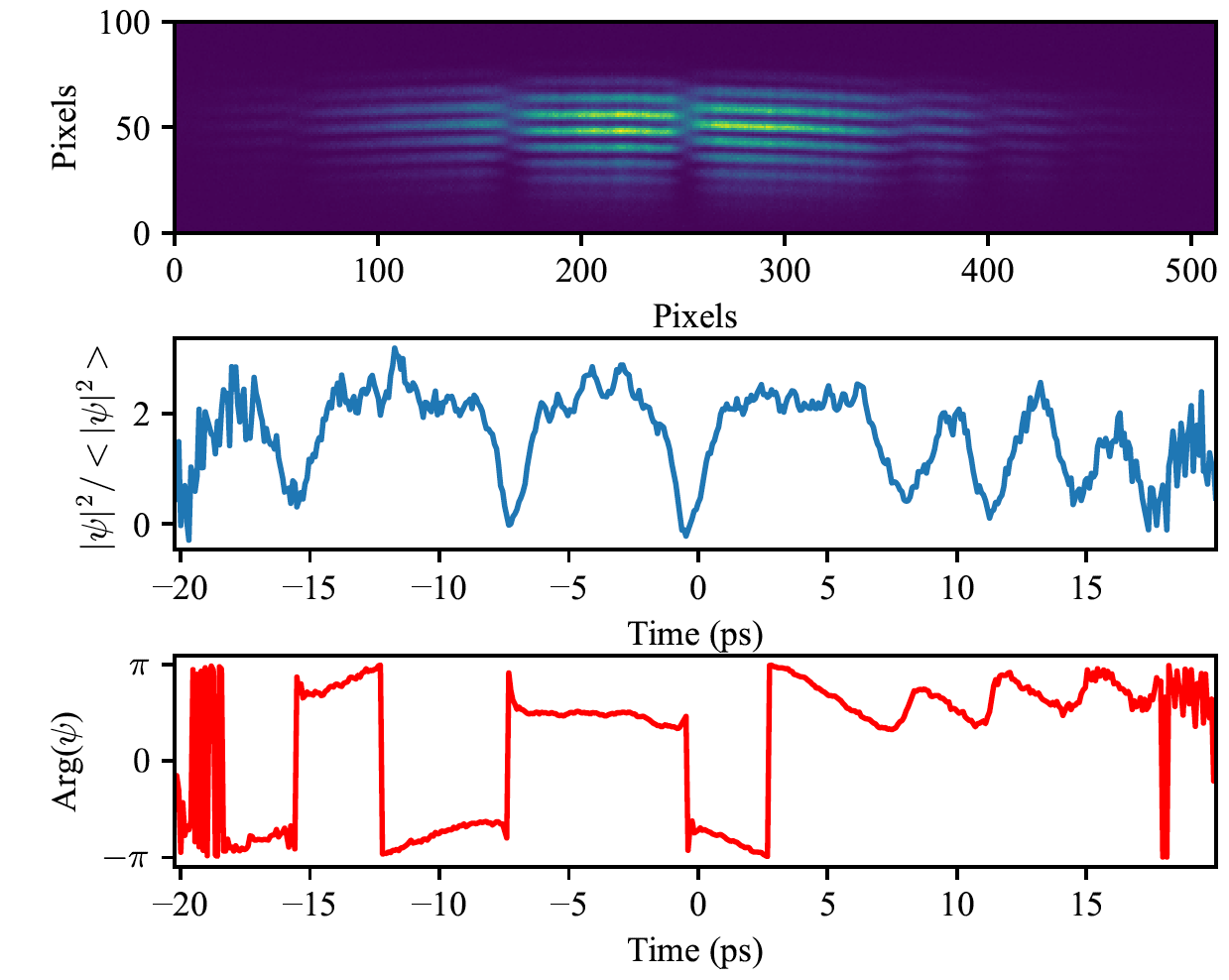}
    \caption{\textbf{An example of the experimentally recorded heterodyne time microscope frame and corresponding power and phase profile extraction.}
(Top) A representative 2-D snapshot recorded experimentally with the heterodyne time microscope, captured by the sCMOS camera. The spatial interference fringes arise from the non-collinear interaction of signal and reference beams at the detection plane, encoding both amplitude (brightness) and phase (relative fringe position) information in the spatial domain. (Middle) Corresponding temporal power profile extracted from the snapshot, obtained by integrating the intensity distribution along the vertical coordinate. (Bottom) Reconstructed temporal phase profile derived through spatial Fourier analysis of the recorded fringe pattern, which contains the relative phase information of the signal beam, with the reference beam’s phase assumed constant across the measurement window.}
	\label{FIG:SI:exp_frame}
\end{figure}

\noindent\textbf{Time Microscope signal description.} The phase and power profiles are reconstructed from the 2-D snapshot (see in Fig.~\ref{FIG:SI:exp_frame}) recorded with the Heterodyne Time Microscope~\cite{tikan2018SingleshotMeasurementPhase,lebel2021SingleshotObservationBreathers}. 
The vertical line \( y \) at position \( x \) of a 2-D snapshot detected with the sCMOS camera of HTM reflects the interference between the non-collinear signal and reference beams resulting in the periodic pattern at a given time. The phase of the reference is consider to be a constant during the snapshot time window. Thus, the relative position of the interference patterns contain the information about the phase of the signal.
The intensity distribution in the 2D snapshot can be formally expressed as:
\begin{multline}
I(x, y) = I_{\mathrm{r}}(x, y) + I_{\mathrm{s}}(x, y) + \\
2 \sqrt{I_{\mathrm{r}}(x, y) I_{\mathrm{s}}(x, y)} \cos \left[k_y y + \phi(x)\right],
\end{multline}
where  \( I_{\mathrm{r}}(x, y) \) transverse reference profile, and \( I_{\mathrm{s}}(x, y) \) - the transverse the signal profile, \( 2\pi / k_y \) is the spatial period of the interference pattern along \( y \), observed by the sCMOS camera, $\phi(x)$ corresponds to the phase of the complex field envelope. 
The angle between the reference and signal beams is chosen to ensure sufficient fringe density, satisfying:
\begin{equation}
\begin{aligned}
\int \sqrt{I_{\mathrm{r}}(x, y) \cdot I_{\mathrm{s}}(x, y)} \cos \left[k_y y + \phi(x)\right] \, \mathrm{d}y \approx 0.
\end{aligned}
\end{equation}

\bigskip
\noindent\textbf{Optical power profile extraction.} 
For each frame, the optical power profile \( P(x) \) of the signal is computed as:
\begin{equation}
\begin{aligned}
P(x) = \int I(x, y) \, \mathrm{d}y - \int I_{\mathrm{r}}(x, y) \, \mathrm{d}y,
\end{aligned}
\end{equation}
where \( P(x) \) is expressed in arbitrary units. The reference intensity is detected in the absence of the signal beam and averaged over 5$\times10^4$ frames to exclude the fluctuations.
The power is normalized as \( P(x) / \langle P(x) \rangle \) or scaled to \( [P(x) / \langle P(x) \rangle] \times P_0 \), where \( P_0 \) is the average power launched inside the fiber. $\langle P(x) \rangle$  is computed with \( 5 \times 10^3 \) frames to exclude the influence of injected power fluctuation during the experiment. We note that since the average power envelope $ \langle P(x) \rangle$ has a limited duration, we observe the degradation of the signal-to-noise ratio at the edge of the extracted signal power \( P(x) / \langle P(x) \rangle \). For this reason, for comparison with the theoretical predictions, we used a window reduced in both $x$ (typically to 200 pixels) and $y$ (100 pixels) directions. Our time calibration measurements gave one pixel in \( x \): \( 79 \, \mathrm{fs} \) correspondence.

\bigskip
\noindent\textbf{Phase profile extraction.} The relative phase \( \phi(x) \) between the signal and reference beams is deduced from the interference fringes. Taking the Fourier transform of the frame in the vertical direction, we extract the phase of the Fourier component that corresponds to the fringes period. Thus the expression for the phase extraction takes the following form:
\begin{equation}
\begin{aligned}
\phi(x) = \text{arg}(\text{max}(\text{FFT}[I(x,y)]_y),
\end{aligned}
\end{equation}
where the maximum amplitude is taken avoiding the low frequency Fourier components.

\bigskip
\noindent\textbf{Computing the probability density function.} To compute the PDF, we analyzed  1000 frames. The confidence interval corresponds to partial averages on four groups equal groups of 250 frames. The data window is restricted to 26 ps in the center of the frame.
We note that the experimental study of the initial conditions and signal after the fiber propagation are recorded separately, in some cases on different days. We, however, ensured the similar performance of the HTM by providing the calibration procedures described in~\cite{tikan2018SingleshotMeasurementPhase}.

\bigskip
\noindent\textbf{Normalization of the NLS.} 
We start with the dimensional form of the NLSE widely used in fiber optics \cite{agrawal2013NonlinearFiberOptics}:
\be
\frac{\partial A}{\partial z_\text{lab}}=-i\frac{\beta_2}{2}\frac{\partial^2 A}{\partial t_\text{lab}^2}+i\gamma |A|^2 A
    \label{eq:nlsexp}\, .
\ee
Where $z_\text{lab}$ and $t_\text{lab}$ are space and time coordinates in the laboratory.
The normalization of the NLS used in the main text is obtained by defining
\be
\psi=\frac{A}{\sqrt{P_0}}\hspace{2pc}
t=-\frac{1}{2}\gamma P_0 \,z_\text{lab}; \hspace{2pc}x=\sqrt{\frac{\gamma P_0}{\beta_2}} \,t_\text{lab}\, ,
    \label{eq:normalization}
\ee
where $\gamma$ is the nonlinear Kerr coefficient, $\beta_2$ is the group velocity dispersion and $P_0$ - the average optical 
power. In these units, the average intensity is normalized to unity $\langle |\psi|^2 \rangle=1$, and the field evolves with the NLS equation
\be
i\partial_t \psi=-\partial_x^2\psi+2|\psi|^2\psi\, .
\ee
The initial random plane wave distribution is described by a gaussian ensemble for the field, most conveniently expressed in the Fourier space $\tilde{\psi}(k)=\int \dd x\, e^{ik x} \psi(x)$. The Fourier components are gaussianly distributed with zero mean $\langle \tilde{\psi}(k)\rangle =0$ and variance $\langle \tilde{\psi}^*(k)\tilde{\psi}(q) \rangle \propto \delta(k-q) e^{-k^2/\Delta k^2}$, the prefactor is determined by the normalization condition $\langle |\psi|^2\rangle=1$ and the width of the distribution is
\be
\Delta k=2\pi\sqrt{\frac{\beta_2}{\gamma P_0}} \,\Delta \nu\, ,
\ee
where $\Delta \nu$ is the experimental signal bandwidth.

\section{Overview of quantum integrability and thermodynamic Bethe ansatz}
\label{sec_qTBA}

In this Appendix, for the sake of completeness, we provide a brief overview on those aspects of the quantum Lieb-Liniger model that are relevant for our scope. A more extensive and in-depth discussion is left to the literature mentioned through the Appendix.

 \bigskip
 \noindent\textbf{The microscopic solution: coordinate Bethe ansatz and Bethe equations.---}
 The determination of the rapidity eigenstates $|\{\lambda_i\}_{i=1}^N\rangle$ is most conveniently framed within the coordinate Bethe ansatz. For a in-depth and pedagogical discussion of this Appendix, see Ref. \cite{Franchini2016}: we consider the Hamiltonian in first quantization within the $N$ particles sector $\hat{H}=\sum_i -\partial_{x_i}^2+c\sum_{i\ne j}\delta(x_i-x_j)$, where $c$ parametrizes the interactions, and look for eigenstates. We consider the bosonic trial wavefunction $\phi_{\{\lambda_i\}_{i=1}^N}(x_1,...,x_N)$ and the ansatz
\be
\label{eq_wave}
\phi_{\{\lambda_i\}_{i=1}^N}(x_1<...<x_N)\propto \sum_{\mathcal{P}}A(\mathcal{P})\prod_{j=1}^N e^{i\lambda_{\mathcal{P}_j} x_j}\, ,
\ee
where the symmetric continuation over all the possible reordering of the coordinates $x_j$ is assumed. The summation is over all the permutations $\mathcal{P}$ of $N$ indexes $\{1,...,N\}$ and $A(\mathcal{P})$ are coefficients to be determined. When the coordinates $x_j$ are all different, the interaction term in the Hamiltonian is ineffective and Eq. \eqref{eq_wave} reduces to a combination of plane waves, which are eigenstate of the derivative part of the Hamiltonian. Requiring  $\phi_{\{\lambda_i\}_{i=1}^N}$ to be an eigenstate also at coincident coordinates leads to constraints on the coefficients $A(\Pi_{j,j+1}\mathcal{P})=S(\lambda_{\mathcal{P}_j}-\lambda_{\mathcal{P}_{j+1}})A(\mathcal{P})$, with $\Pi_{j,j+1}$ the permutation of the indexes $j$ and $j+1$, and $S(\lambda)=\tfrac{\lambda+ic}{\lambda-ic}$ the so-called scattering matrix: in the LL model, $S(\lambda)$ is just a complex number with modulus one. An explicit solution for $A(\mathcal{P})$  can be found, but it is not crucial for our scope, and $\phi_{\{\lambda_i\}_{i=1}^N}$ \eqref{eq_wave} becomes a true eigenvector of the Hamiltonian and of all the extensive conserved charges $\hat{Q}_n |\{\lambda_i\}_{i=1}^N\rangle=\left(\sum_{j=1}^N q_n(\lambda_j)\right)|\{\lambda_i\}_{i=1}^N\rangle$. 
The charge eigenvalue $q(\lambda)$ associated to the Hamiltonian is $\lambda^2$, while for the momentum and number of particles one has $\lambda$ and $1$ respectively. In general, charge eigenvalues of the form $q_n(\lambda)=\lambda^n$ are associated to the extensive conservation laws.

When considering thermodynamics, it is crucial having a finite density of particles. We assume the quantum particles live on a interval of length $L$ and ask for periodic boundary conditions (PBC) for the wavefunction: other boundary conditions, for example open boundary conditions, lead to the same thermodynamics when $L\to \infty$. Imposing PBC on Eq. \eqref{eq_wave} leads to the Bethe equations \cite{Bethe1931}, a set of nonlinear equations that quantize the rapidities
\be
e^{iL\lambda_j }=\prod_{k\ne j}S(\lambda_j-\lambda_k)\hspace{2pc}j,k\in \{1,...,N\}\, .
\ee
All the solutions to the Bethe equations are associated to physical eigenstates Eq. \eqref{eq_wave}, provided the rapidities are chosen all different $\lambda_j\ne \lambda_k$: in case of equal rapidities, the wavefunction vanishes and so those configurations may be excluded. In the thermodynamic limit, complex solutions to the Bethe equations are associated with bound states and their presence depends on the sign of the interactions: for the repulsive case $c>0$ we are interested in, only real solutions are important. 

\bigskip
\noindent\textbf{GGEs and rudiments of thermodynamic Bethe ansatz.---}
Armed with the Bethe equations, one can forget about the explicit form of the eigenstates \eqref{eq_wave} in building the thermodynamics within the thermodynamic Bethe anstaz (TBA) \cite{takahashi2005thermodynamics}, which we briefly overview (see also Ref. \cite{Franchini2016}).
First, one considers the Bethe equations in logarithmic form and parametrizes the rapidities in terms of Bethe integers $I_j\in \mathbb{Z}$ as 
\be\label{eq_logBE}
\frac{I_j}{L}=\frac{\lambda_j}{2\pi}-\frac{1}{2\pi L}\sum_{k\ne j}\Theta_\text{q}(\lambda_j-\lambda_k)\, ,
\ee
where one defines the scattering phase as $\Theta_\text{q}(\lambda)\equiv -i \log S(\lambda)$: the eigenstates $|\{\lambda_i\}_{i=1}^N\rangle$ are now in one-to-one correspondence with all the set of distinct integers $\{I_j\}_{j=1}^N$.
Let us now consider the GGE partition function $\mathcal{Z}=\text{Tr}[e^{-\sum_n \beta_n \hat{Q}_n}]$, which one rewrites as an explicit sum over the eigenstates parametrized either with the rapidities, or the Bethe integers. The second choice is more convenient for further manipulations $\mathcal{Z}=\sum_N\sum_{\{I_j\}_{j=1}^N} e^{-\sum_{j=1}^N\sum_n\beta_n q_n(\lambda_j)}$, where $I_j$ and $\lambda_j$ are connected through the Bethe equations.
We are now ready to move to the thermodynamic limit and, upon defining $y_j=I_j/L$, introduce a counting function $\sigma(y)$ which counts the number of $y_j$ within an interval of size $\dd y$, i.e. $L\dd y\sigma(y)=\{\# y_j\in [y-\tfrac{\dd y}{2},y+\tfrac{\dd y}{2}]\}$. The sum in the definition of the partition function is now readily converted into a path integral over the possible choices of the coarse grain counting function $\sigma$. When doing so, one should introduce a proper entropic weight due to the possible microscopic rearrangement of $y_j$ that do not affect $\sigma(y)$. Since $y_j$ behaves as hard-core particles, this eventually results in a Fermi-Dirac statistics \cite{Franchini2016}
\be
\label{eq_ZI}
\mathcal{Z}\simeq \int \mathcal{D} \sigma\, \exp\Bigg[L\int \dd y  f(\sigma(y))-
L\int \dd\lambda \sum_n \beta_n q(\lambda(y))\sigma(y)\Bigg]
\ee
where $ f(x)=-x\log x-(1-x)\log(1-x)$ and one introduces the $\lambda(y)$ function from the coarse grain limit of the logarithmic Bethe equations $y=\frac{\lambda(y)}{2\pi}-\frac{1}{2\pi }\int \dd y'\Theta_\text{q}(\lambda(y)-\lambda(y'))\sigma(y')$. As a last step, in Eq. \eqref{eq_ZI} it is convenient to change variable $y\to \lambda$ by introducing the total root density $\rho^t(\lambda)$ defined through the Jacobian of the transformation $\rhoq^t(\lambda)\equiv \tfrac{\dd y}{\dd \lambda}$, and the root density $\rhoq(\lambda)=\rhoq^t(\lambda) \sigma(y(\lambda))$. After this change of variables, the GGE partition function is rewritten as
\be\label{eq_Zpath}
\mathcal{Z}\simeq \int \mathcal{D} \rho\, \exp\left[L \mathcal{S}[\rhoq]-L\int \dd\lambda \left(\sum_n q_n(\lambda)\right)\rhoq(\lambda)\right]\, ,
\ee
where we define the entropy density 
\be\label{eq_Scl}
\mathcal{S}[\rhoq]=\int \dd\lambda \, \rhoq^t f(\vartheta_\text{q}(\lambda))\, ,
\ee
and the filling fraction $\vartheta_\text{q}(\lambda)\equiv \rhoq(\lambda)/\rhoq^t(\lambda)$. Computing the path integral \eqref{eq_Zpath} is a dauting task, but in the thermodynamic limit $L\to +\infty$ only the saddle point matters. 
Looking for the saddle point finally leads to integral equations connecting the Lagrange multipliers and the filling function (and thus the root density)
\begin{multline}\label{eq_TBA}
f'(\vartheta(\lambda))=\sum_n \beta_n q_n(\lambda)+\\
-\int \frac{\dd\lambda'}{2\pi}\varphi_\text{q}(\lambda-\lambda')[\vartheta_\text{q}(\lambda)f'(\vartheta_\text{q}(\lambda))-f(\vartheta_\text{q}(\lambda))]\, .
\end{multline}
Above, the scattering kernel is defined as $\varphi_\text{q}(\lambda)\equiv\partial_\lambda \Theta_\text{q}(\lambda)$, and in Lieb Liniger it reads $\varphi_\text{q}(\lambda)=-\frac{2c}{c^2+\lambda^2}$ \cite{Lieb1963A}.

\bigskip
\noindent\textbf{The density moments.---} The TBA equation \eqref{eq_TBA} is a cornerstone in describing thermodynamics and more in general GGEs, and it already allows for a direct computation of certain observables. For example, apart from more naive expectation values of local charges $\langle \hat{Q}_n\rangle=L\int \dd\lambda q_n(\lambda)\rho(\lambda)$, their connected correlators can be computed through derivatives of the partition function $\langle \hat{Q}_{n}\hat{Q}_{n'}\rangle_\text{conn.}=-\tfrac{\partial^2}{\partial \beta_n \partial \beta_{n'}}\log \mathcal{Z}$. In principle, the knowledge of $\rho(\lambda)$ and, through \eqref{eq_TBA}, of the whole GGE determines the expectation values of all local observables. However, obtaining analytical formulas is a formidable task.
Recently, close integral expressions for the density moments $\langle (\hat{\psi}^\dagger)^n(\hat{\psi}^\dagger)^n\rangle$ have been computed in Ref. \cite{Bastianello2018,Bastianello2018B}, and are encoded in the following generating function 
\be\label{eq_q_gen}
1+\sum_{n=1}^\infty Y^n\frac{(2c)^n}{n!}\langle (\hat{\psi}^\dagger)^n(\hat{\psi}^\dagger)\rangle=\exp\left(\frac{1}{\pi} \sum_{n=1}^\infty Y^n \mathcal{G}_n\right)\, ,
\ee
from which the moments can be obtained upon expanding both sides in powers of the auxiliary variable $Y$. Above, $\mathcal{G}_n=\frac{c^{2n-1}}{n}\int \dd\lambda\, \vartheta_{\text{q}}(\lambda)\xi_{2n-1}^\dr(\lambda)$, where the dressing operation $\xi_{j}(\lambda)\to \xi_{j}^\dr(\lambda)$ is in general defined as the solution of the following integral equation
\be
\xi^\dr_{j}(\lambda)=\xi_{j}(\lambda)-\int \frac{\dd\lambda'}{2\pi}\varphi_\text{q}(\lambda-\lambda')\vartheta_\text{q}(\lambda')\xi^\dr_{j}(\lambda')\, ,
\ee
and the auxiliary functions $\xi_{j}(\lambda)$ are determined by recursive integral equations
\begin{eqnarray}
\nonumber &&\xi_{2n}(\lambda)=\int \frac{\dd\lambda}{2\pi}\vartheta_\text{q}(\lambda)\big\{\Gamma(\lambda-\lambda')[2 \xi_{2n-1}^\dr(\lambda')-\xi_{2n-3}^\dr(\lambda')]+\\
\nonumber&&\varphi_\text{q}(\lambda-\lambda') \xi_{2n-2}^\dr(\lambda')\big\}\, ,\\
\nonumber&& \xi_{2n+1}(\lambda)=\delta_{n,0}+\int \frac{\dd\lambda'}{2\pi}\vartheta_\text{q}(\lambda')\vartheta_\text{q}(\lambda')\big\{\Gamma(\lambda-\lambda')\xi_{2n}^\dr(\lambda')+\\
&&\varphi(\lambda-\lambda')\xi_{2n-1}^\dr(\lambda')\big\}\, ,
\end{eqnarray}
with $\Gamma(\lambda)=-\frac{\lambda}{c}\varphi(\lambda)$ and $\xi_{n\le 0}(\lambda)=0$. These recursive equations can be efficiently solved numerically \cite{Bastianello2018B}.
This result on the density moments on a GGE in the quantum model gives access to the classical PDF through semiclassical limits, see Appendix \ref{sec_qtocl}.

\section{Overview of classical integrability and thermodynamic Bethe ansatz}
\label{sec_clTBA}

In this Appendix, we provide a short overview of the classical inverse scattering for the defocusing NLS and its connection with thermodynamics. 

\bigskip
\noindent\textbf{Lax pairs and monodromy matrix.--} The NLS equation emerges as the compatibility condition for an auxiliary linear problem \cite{Faddeev1987}. Let us define the $\lambda-$dependent Lax pair
\begin{eqnarray}
&&U_\lambda(t,x)=(\psi^*\sigma^++\psi \sigma^-)+\tfrac{\lambda}{2i}\sigma^z\, ,\\
\nonumber&&V_\lambda(t,x)=\left(i|\psi|^2+\frac{i\lambda^2}{2}\right)\sigma^z-[\lambda \sigma^x+i\partial_x\psi^*\sigma^+-i\partial_x\psi\sigma^-]\, ,
\end{eqnarray} 
with $\sigma^\pm=(\sigma^x\pm \sigma^y)/2$, and introduce the auxiliary field $\Psi(t,x)\in \mathbb{C}^2$ obeying the following linear differential equation
\be\label{eq_aux}
\partial_x \Psi(t,x)=U_\lambda(t,x) \Psi(t,x)\, ,\hspace{1pc}\partial_t \Psi(t,x)=V_\lambda(t,x)\Psi(t,x)\, .
\ee
The consistency of the auxiliary problem \eqref{eq_aux} requires the zero-curvature condition $\partial_\lambda U_\lambda-\partial_x V_\lambda+[U_\lambda,V_\lambda]=0$ which is satisfied if and only if the field $\psi$ obeys the NLS equation. Describing the NLS as a compatibility condition for a Lax pair manifests integrability and the presence of conservation laws.
Below, we need to discuss separately two different boundary conditions for the NLS field: (i) the ``whole line problem" where the field $\psi(t,x)$ vanishes at $|x|\to \infty$ faster than any power law, and (ii) the periodic problem $\psi(t,x)=\psi(t,x+L)$. A thermodynamic description of the state can be achieved combining these two problems.

\bigskip
\noindent\textbf{The whole-line problem.---} Let us define the spatial propagator for the auxiliary problem \eqref{eq_aux} $T(t,x,y)=\text{Pexp}\left[\int_x^y \dd z\, U_\lambda(t,z)\right]$, which solves the simple equation $\partial_t T_\lambda(t,x,y)=V_\lambda(t,x)T_\lambda(t,x,y)-T_\lambda(t,x,y)V_\lambda(t,y)$. It is immediate to check that, whenever the interval $(x,y)$ is chosen sufficiently large in such a way the interval contains the whole non-trivial part of the field $\psi$, then $\partial_t \text{Tr}[T_\lambda(t,x,y)]=0$ for any $\lambda$. Since the field $\psi$ initially vanishing outside of the interval $(x',y')$ can propagate beyond the initial interval during time evolution, we now send the extrema to infinity. However, for practical purposes it should be remembered that the considerations below apply for a finite interval as well, as long as the field vanishes out of it.
We then define the transfer matrix on the whole line as \cite{Faddeev1987,DeLuca2016}
\be\label{eq_transfer}
\mathcal{T}_\lambda(t)=\lim_{x\to\infty}  e^{\tfrac{-\lambda x}{2i}\sigma^z}T_\lambda(t,-x,x)e^{-\tfrac{\lambda x}{2i}\sigma^z}\, ,
\ee
where the additional oscillating terms counterbalance the asymptotic behavior of $U_\lambda(t,x)$. Based on the symmetries of the linear problem only, the transfer matrix can be parametrized as
\be\label{eq_Tab}
\mathcal{T}_\lambda(t)=\begin{pmatrix}a(\lambda) & b^*(\lambda) \\ b(\lambda) & a^*(\lambda)\end{pmatrix}\, ,
\ee
where for the sake of notation, the time-dependence of the matrix elements is left implicit. We have already seen as $\partial_t \text{Tr}[\mathcal{T}]_\lambda(t)=\partial_t [a(\lambda)+a^*(\lambda)]=0$, but for vanishing boundary conditions it holds the stronger statement $\partial_t a(\lambda)=0$, as it can be immediately checked from the equation for the time evolution of the propagator $T_\lambda(t,x,y)$.
The coefficient $a(\lambda)$ is conserved for every $\lambda$, and it can be esplictly shown \cite{Faddeev1987} the Laurent expansion of its logarithm generates the extensive charges $\log a(\lambda)\stackrel{\lambda\to\infty}{=}\sum_n \frac{1}{\lambda^{n+1}}\mathcal{Q}_n$. We will now use this fact and the analiticity properties of $a(\lambda)$ to recover the classical counterpart of the quantum root density: an in depth discussion is left to the literature \cite{Faddeev1987,DeLuca2016}, here we summarize the main facts.
The coefficient $a(\lambda)$ can be analytically continued in the complex plane and $
\lim_{|\lambda|\to \infty}a(\lambda)\to 1$: in the focusing regime, $a(
\lambda)$ can have zeroes and poles in the complex plane, which are associated with solitons, while these are absent in the defocusing regime of interest for us. Therefore, one can use the Kramers-Kronig relations and parametrize $a(\lambda)$ as
\be\label{eq_KK}
a(\lambda)=\exp\left[\int \frac{\dd\lambda'}{\pi i} \frac{\log|a(\lambda')|}{\lambda'-\lambda-i 0^+}\right]\, .
\ee
By taking the logarithm, and considering the Laurent expansion one reaches the identification $\mathcal{Q}_n=\int \dd\lambda\, \lambda^n \tfrac{\log|a(\lambda)|}{\pi}$. 
A final identification of $\log|a(\lambda)|$ with the classical root density passes through the discussion of the periodic problem.

\bigskip
\noindent\textbf{The periodic problem.---} We here assume that the NLS field is periodic $\psi(t,x+L)=\psi(t,x)$, for further details see \cite{Forest1982,DeLuca2016}. As the auxiliary problem is linear and thus symmetric under rescaling of $\Psi(t,x)$, one does not ask the auxiliary field to be periodic, but allows for more freedom allowing periodicity modulus an overall complex constant $\Psi(t,x)=A\Psi(t,x+L)$ with unitary norm $|A|=1$. A different choice $|A|\ne 1$ would lead to not normalizable solutions for $\Psi$ in the thermodynamic limit, and are thus excluded. As we see below, $A$ is further constrained. 
We introduce the PBC transfer matrix from the propagator as $\mathcal{T}_\lambda^\text{PBC}=T_\lambda(t,0,L)$: notice that, in contrast with Eq. \eqref{eq_transfer}, further oscillating terms are not included. In the periodic problem, the transfer matrix can still be parametrized as in Eq. \eqref{eq_Tab}, and therefore  $[\mathcal{T}_\lambda]_{11}=[\mathcal{T}_\lambda]_{22}^*$: this leaves as only possibility for the boundary conditions $A=\pm 1$.
Notice that while the trace of $\mathcal{T}_\lambda^\text{PBC}$ is still a conserved quantity, the diagonal entries are not separately conserved.
The periodic boundary conditions $\mathcal{T}_\lambda^\text{PBC}=\pm \text{Id}$ can be fulfilled only for discrete values of the spectral parameter $\lambda$, and leads to the classical counterpart of the Bethe equations introduced in Appendix \ref{sec_qTBA}.

\bigskip
\noindent\textbf{The emergence of thermodynamics.---} We can now combine the knowledge of the whole-line and periodic problem to take the thermodynamic limit, see Ref. \cite{DeLuca2016} for an extensive discussion. We consider a fundamental interval $\mathcal{I}_1=[-L/2,L/2]$, and a smaller one $\mathcal{I}_2\subset \mathcal{I}_1$ defined as $\mathcal{I}_2=[-L/2+\ell/2,L/2-\ell/2]$. Eventually, we will be interested in the limit when both $L$ and $\ell$ are sent to infinity, but $\ell/L\to 0$. We consider smooth field configurations $\psi$ with non-vanishing support in $\mathcal{I}_2$ and zero in the complement of $\mathcal{I}_1$: these field configurations can be seen alternatively as belonging either to the whole line problem by continuing the field as zero to infinity, or as the PBC problem by considering periodic repetitions of the fundamental interval. It is useful to conveniently switch between the two viewpoints.
As long as the time evolution does not propagate the NLS field outside of $\mathcal{I}_1$, we can compare the whole-line transfer matrix \eqref{eq_transfer} with the periodic one, and in particular have $a(\lambda)=e^{i L\lambda/2}a^\text{PBC}(\lambda)$: boundary conditions on the PBC transfer matrix impose $a^\text{PBC}(\lambda_j)=\pm 1$, by taking the logarithm and using the representation \eqref{eq_KK}, one reaches \cite{DeLuca2016}
\be\label{eq_clBE}
\frac{ I_j}{L} = \frac{\lambda_j}{2\pi} -\fint \frac{\dd\lambda'}{2\pi } \frac{2}{\lambda_j-\lambda'}\frac{\log|a(\lambda')|}{L \pi}\, .
\ee
The similarity with the logarithmic form of the Bethe equations \eqref{eq_logBE} is now evident, and this analogy, together with the fact that $\log|a(\lambda)|$ generates the conserved quantities, further encourages to identify the classical root density as $\rho(\lambda)\equiv \lim_{L\to \infty} \tfrac{1}{\pi}\log|a(\lambda)|$ \cite{DeLuca2016,DelVecchio2020,Bezzaz2023}, and the \emph{classical scattering phase} $\Theta(\lambda)=\tfrac{2}{\lambda}$, then leading to the classical scattering kernel $\varphi(\lambda)=\partial_\lambda \Theta(\lambda)$. The argument presented here is not sufficient to derive the thermodynamics by paralleling the framework of TBA discussed in Appendix \ref{sec_qTBA}, as Eq. \eqref{eq_clBE} is already in a coarse grained form and hinders the possibility of counting the microscopic solutions and characterize the classical entropy. A rigorous derivation passes through the identification of the action-angle microscopic variables \cite{Bullough1986}, leading to the microscopic analogue of Eq. \eqref{eq_clBE}. We will not go through this highly technical detour, as we provide an alternative derivation by considering the semiclassical limit of the LL in Appendix \ref{sec_qtocl}. The final outcome is an entropy of the general form \eqref{eq_Scl}, provided the classical root density is used and that the entropy factor is of Rayleigh-Jeans type $f(x)=\log x$.

\section{From Lieb-Liniger to Non Linear Schroedinger: semiclassical limits and PDF}
\label{sec_qtocl}

For completeness, we provide a short summary of the semiclassical limit connecting the Lieb-Liniger model and the NLS, in the defocusing regime (see Ref. \cite{Koch2022} for the focusing case). This Appendix summarizes the material presented in Ref. \cite{DelVecchio2020}, to which we refer for a more in-depth discussion.

\bigskip
\noindent\textbf{The semiclassical limit of thermodynamics.---} The semiclassical limit is achieved for weak interactions and in the large occupation limit: in short, the energy scale should be such that the quantum model cannot resolve the addition of a single excitation, which can be then coarse grained to a continuous variable. The semiclassical limit does not require integrability, and it is best introduced by looking at thermal expectation values.
We consider a thermal partition function $\mathcal{Z}=\text{Tr} e^{-\beta_\text{q}(\hat{H}-\mu \hat{N})}$ with $\hat{N}$ the number of particles, and introduce a path integral representation of the quantum field $\hat{\psi}(x)\to \phi(\tau,x)$ with the Euclidean time having compact support $\tau\in [0,\beta_\text{q}]$, and conveniently rewrite the field in terms of the modes in the Euclidean time $\phi(\tau,x)=\sum_{n=-\infty}^\infty \frac{e^{i 2\pi n\tau/\beta_\text{q}}}{\sqrt{c}}\psi_n(x)$. 
The semiclassical limit is achieved in a double limit of weak interaction $c\to 0$ and large temperature $\beta_\text{q}\to0$, in such a way $\beta=\beta_\text{q}/c$, with $\beta$ the classical temperature.
In terms of the partition function one has
\begin{multline}
\mathcal{Z}=\int \mathcal{D}\phi \,e^{-\int \dd x\int_0^{\beta_\text{q}} \dd\tau \frac{1}{2}(\phi^*\partial_\tau \phi-\phi\partial_\tau \phi^*) +|\partial_x\phi|^2+c|\phi|^4+\mu |\phi|^2}=\\
\int \prod_n \mathcal{D}\psi_n\, \exp\Bigg\{-\frac{\beta_\text{q}}{c} \int \dd x\, \sum_n\left[ \left[\left(\tfrac{2\pi n }{\beta_\text{q}}\right)^2+\mu\right]|\psi_n|^2+|\partial_x\psi_n|^2\right]+\\
\sum_{n_1+n_2=n_3+n_4}\psi_{n_1}^*\psi_{n_2}^*\psi_{n_3}\psi_{n_4}\Bigg\}\, .
\end{multline}
The double limit is immediately understood, as for small $\beta_\text{q}$ all the oscillating modes $\psi_{n\ne0}$ are largely suppressed. The only non-vanishing field is the zero mode $\psi_{n=0}(x)$ which can be identified with the classical field of the NLS $\psi_{n=0}(x)\to \psi(x)$, and the quantum partition function becomes the classical partition function. This analysis can be extended to time evolution, and also to the integrability framework and GGEs with the correspondence $\beta_{\text{q},n}\simeq c \beta_n$. The mode decomposition over the Matsubara frequencies also establishes the correspondence between quantum and classical fields $\hat{\psi}(x)\simeq \frac{1}{\sqrt{c}}\psi(x)$, which holds in a weak sense when computing correlation functions.

\bigskip
\noindent\textbf{The semiclassical limit of the TBA and PDF.---} The semiclassical limit can be taken at the level of thermodynamic Bethe ansatz, and the generating function for the moments of the density \eqref{eq_q_gen} can be also projected to the classical limit finally leading to the formula for the PDF presented in the main text. We leave the detailed calculations to the Ref. \cite{DeLuca2016,DelVecchio2020}, and we limit ourselves to summarize the key-points.
The correspondence at the leading order in the semiclassical limit between the quantum root density $\rhoq(\lambda)$, total root density $\rhoq^t(\lambda)$, and scattering kernel $\varphi_\text{q}(\lambda)$ with the classical counterparts is
\begin{eqnarray}
\nonumber&&\rhoq(\lambda)\simeq \frac{1}{c}\rho(\lambda)\\
\nonumber&&\rhoq^t(\lambda)-\rhoq(\lambda)\simeq \rho^t(\lambda)\\
&&
\varphi_\text{q}(\lambda)\simeq \delta(\lambda)+c\varphi(\lambda)
\end{eqnarray}

All the general formulas of the quantum TBA becomes the analogue expression in the classical variables upon the above substitution. Furthermore, the semiclassical limit of the quantum entropy \eqref{eq_Scl} becomes the classical entropy discussed at the end of Appendix \ref{sec_clTBA}. The minimization of the free energy leads to the same integral equations \eqref{eq_TBA}, provided the Rayleigh-Jeans distribution $f(x)=\log x$ is used.
As discussed in Ref. \cite{DelVecchio2020}, the semiclassical limit can be also extended to the formula for generating moments of $|\psi|^2$ \eqref{eq_q_gen}. However, this does not give directly the PDF: in Ref. \cite{DelVecchio2020} an equivalent representation of the moment generating function is used, which is easier to be inverted to finally obtain the sought formula for the PDF
\be\label{eq_P}
P_\text{GGE}(|\psi|^2=d)=16\int_0^\infty \dd \theta J_0(8\sqrt{\theta d})e^{-\frac{16}{\pi}\int_0^{\sqrt{\theta}}\dd\tau\, \tau S(\tau)}\, ,
\ee
with $J_0(x)$  a modified Bessel function of the first kind, the classical filling and total root density are $\vartheta(\lambda)=\rho(\lambda)/\rho^t(\lambda)$ and $\rho^t(\lambda)=\tfrac{1}{2\pi}-\fint\frac{\dd\lambda'}{2\pi}\tfrac{2}{\lambda-\lambda'}\partial_{\lambda'}\rho(\lambda')$, and $S(\tau)\equiv \int \dd\lambda s_{\tau}(\lambda)$.
The function $s_{\tau}(\lambda)$ is the solution of the auxiliary problem
\be\label{eq_sSM}
[\vartheta^{-1}(\lambda)+i0^+]s_\tau(\lambda)=1+\fint \frac{\dd\lambda'}{2\pi} \frac{2}{\lambda-\lambda'}(4i\tau-\partial_{\lambda'})s_\tau(\lambda')\, .
\ee
The detailed derivation is left to the original reference \cite{DelVecchio2020}.
Here, we provide a quick consistency check verifying that, in the limit of weak non-linearity, $P_\text{GGE}$.
We achieve weak non-linearities in the limit of a broad plane-wave momentum distribution in the initial conditions $\Delta k\to \infty$.
In this limit, the root density converges to the momentum distribution and we can therefore approximate it $\rho(\lambda)\simeq\frac{1}{\Delta k \sqrt{\pi}}e^{-\lambda^2/\Delta k^2}$ and, at the leading order, $\rho^t(\lambda)\simeq \tfrac{1}{2\pi}$, resulting in $\vartheta(\lambda)\simeq  \frac{2\sqrt{\pi}}{\Delta k}e^{-\lambda^2/\Delta k^2}$. We can now use this approximation in Eq. \eqref{eq_sSM} and notice that, since $\vartheta(\lambda)$ is very small, at the leading order in $\Delta k$ the solution for Eq. \eqref{eq_sSM} is $s_\tau(\lambda)\simeq \vartheta(\lambda)$, leading to $S(\tau)\simeq \frac{1}{2\pi}$. Importantly $S(\tau)$ is $\tau-$independent at this order. Using the approximate value of $S(\tau)$ in Eq. \eqref{eq_P} we obtain
\be
P_\text{GGE}(|\psi|^2=d)\simeq 16\int_0^\infty \dd \theta J_0(8\sqrt{\theta d})e^{-16\, \theta }\, ,
\ee
This last integral can be performed with Mathematica resulting in $P_\text{GGE}(|\psi|^2=d)\simeq e^{-d}$, as expected in the limit of zero non-linearity.

\bigskip
\noindent\textbf{How to compute the PDF in practice.---} We briefly discuss how to numerically solve the formula for the PDF: a working code is provided on Zenodo \cite{Zenodo}.
The TBA equations for computing the total root density and the auxiliary function $s_\tau(\lambda)$ can be discretized with standard methods and transformed in matrix-valued equations, see eg. \cite{DelVecchio2020} and \cite{Zenodo}.
The crucial point is correctly computing $e^{-\frac{16}{\pi}\int_0^{\sqrt{\theta}}\dd\tau S(\tau)}$: this function rapidly vanishes at large $\theta$ and thus allows for a truncation of the integral in Eq. \eqref{eq_P}, but it develops zeroes associated with poles in $S(\tau)$. These poles need to be extracted and properly accounted for. We first properly rewrite \eqref{eq_sSM} in an operatorial form $\int\dd\lambda'[\Omega_\tau(\lambda,\lambda')+i0^+\delta(\lambda-\lambda')]s_\tau(\lambda')=1$, the definition of $\Omega_\tau(\lambda,\lambda')$ is readily obtained by a direct comparison with Eq. \eqref{eq_sSM}. $\Omega_\tau(\lambda,\lambda')$ is parametrically dependent on $\tau$, and as an operator on the $\lambda-$space is Hermitian, and thus can be diagonalized. Let us call $b_{n;\tau}(\lambda)$ the eigenfunctions with eigenvalues $\mu_n(\tau)$, we assume $b_{n;\tau}(\lambda)$ are normalized to unity $\int \dd\lambda\, |b_{n;\tau}(\lambda)|^2$.
In this representation, we can rewrite $S(\tau)$ as
\be
S(\tau)=\sum_n \frac{B_n(\tau)}{\mu_n(\tau)+i0^+}\, ,\hspace{2pc} B_n\equiv \int \dd\lambda \, b_{n;\tau}(\lambda)\,.
\ee
One needs now to isolate the possible singularities. Let us fix a maximum cutoff $\tau<\sqrt{\theta_\text{max}}$ for the computation of the PDF \eqref{eq_P}. We tabulate the spectrum of the operator $\Omega_\tau(\lambda,\lambda')$ scanning through $\tau\in[0,\sqrt{\theta_\text{max}}]$ and looking for solutions $\mu_n(\tau)=0$, which are then tabulated as pairs $(n_j,\tau_j)$ in such a way $\mu_{n_j}(\tau_j)=0$. We split $S(\tau)$ in a regular and singular part
\be
\tau S(\tau)=\tau S^\text{reg}(\tau)+\sum_j\frac{\tau_jB_{n_j}(\tau_j)}{\partial_\tau \mu_{n_j}(\tau_j)(\tau-\tau_j)+i0^+}\, .
\ee
As the next step, we can explicitly integrate over the poles when computing $\int \dd\tau \tau S(\tau)$. The explicit integration of one of these poles gives
\begin{widetext}
\be
\exp\left[-\frac{16}{\pi}\int_0^{\sqrt{\theta}}\dd\tau \, \tau \frac{\tau_jB_{n_j}(\tau_j)}{\partial_\tau \mu_{n_j}(\tau_j)(\tau-\tau_j)+i0^+}\right]=\left|\frac{\sqrt{\theta}-\tau_j}{\tau_j}\right|^{-\frac{16}{\pi}\frac{\tau_j B_{n_j}(\tau_j)}{\partial_\tau \mu_n(\tau_j)}} \exp\left[\frac{16}{\pi}\frac{\tau_j B_{n_j}(\tau_j)}{|\partial_\tau \mu_n(\tau_j)|} i\pi \text{H}(\sqrt{\theta}-\tau_j)\right]\, ,
\ee
\end{widetext}
with $\text{H}(x)$ the Heaviside Theta function $\text{H}(x>0)=1$ and zero otherwise. Apparently, the function above is non-analytic: however, it can be analytically shown \cite{Bullough1986} that $-\frac{16}{\pi}\frac{\tau_j B_{n_j}(\tau_j)}{\partial_\tau \mu_n(\tau_j)}=1$, leading to simple zeroes $\exp\left[-\frac{16}{\pi}\int_0^{\sqrt{\theta}}\dd\tau \, \tau \frac{\tau_jB_{n_j}(\tau_j)}{\partial_\tau \mu_{n_j}(\tau_j)(\tau-\tau_j)+i0^+}\right]=\tfrac{\sqrt{\theta}-\tau_j}{\tau_j}$. Notice that in numerical discretizations of the kernel $\Omega_\tau(\lambda,\lambda')$, the identity $-\frac{16}{\pi}\frac{\tau_j B_{n_j}(\tau_j)}{\partial_\tau \mu_n(\tau_j)}=1$ will not be exact, but will have small corrections due to the discretization. After having explicitly integrated over the singular terms, one is left with
\be
e^{-\frac{16}{\pi}\int_0^{\sqrt{\theta}}\dd\tau\, \tau S(\tau)}=e^{-\frac{16}{\pi}\int_0^{\sqrt{\theta}}\dd\tau\, \tau S^\text{reg}(\tau)}\prod_j\left(\frac{\sqrt{\theta}-\tau_j}{\tau_j}\right)\, ,
\ee
This is the most convenient expression for a numerical implementation, where in the first step one finds the zeroes in the spectrum of $\Omega_\tau$ finding $\tau_j$ and removing the singularities from $S(\tau)$, then the integral over the regular part of $S^\text{reg}(\tau)$ can be safely evaluated. This implementation is used in the code provided on Zenodo \cite{Zenodo}.

\newpage

\bibliography{biblio}

\end{document}